\documentclass[seceq,supplement]{ptptex}

\usepackage{graphicx}
\usepackage{wrapfig}

\def\bi#1{\mbox{\boldmath$#1$}}

\title{ Reproducibility of a noisy limit-cycle oscillator\\
  induced by a fluctuating input }

\author{% %Use \scshape for the family name
  Hiroya {\scshape Nakao},
  Ken {\scshape Nagai},
  and
  Ken-suke {\scshape Arai}
}

\inst{
  Department of Physics, Kyoto University, Kyoto 606-8502, Japan
}

\recdate{\today}

\abst{ Reproducibility of a noisy limit-cycle oscillator driven by a
  random piecewise constant signal is analyzed. By reducing the model
  to random phase maps, it is shown that the reproducibility of the
  limit cycle generally improves when the phase maps are monotonically
  increasing.  }

\begin{document}

\maketitle

\section{Introduction}

When a spiking neuron receives a randomly fluctuating input, its
reproducibility of spike generation improves compared with the case of
a constant input~\cite{Mainen}.
This phenomenon can be interpreted as phase synchronization between
uncoupled nonlinear oscillators that receive a common fluctuating
input, because repeated measurements on a single oscillator using the
same input is equivalent to a single measurement on an ensemble of
uncoupled identical oscillators.
In our previous studies, we analyzed the cases where the fluctuating
input is given by a random telegraphic signal~\cite{Nagai} or by a
random impulsive signal~\cite{Nakao}.
In this proceeding, we analyze the case where the fluctuating input is
a slowly varying, piecewise constant random signal using the phase
reduction technique~\cite{Kuramoto,Winfree}, as a generalization
towards a full treatment of realistic continuous random signals.

\section{Fluctuation-induced phase synchronization}

We consider an ensemble of $N$ identical uncoupled limit-cycle
oscillators subject to a common fluctuating input:
\begin{equation}
  {\dot {\bi X}_i}(t) = {\bi G}({\bi X}_i(t)) + {\bi I}(t)
  \label{Eq:ode}
\end{equation}
for $i = 1, \cdots, N$, where ${\bi X}_i(t)$ represents the internal
state of the $i$-th oscillator at time $t$, ${\bi G}({\bi X})$ the
intrinsic dynamics of each oscillator, and ${\bi I}(t)$ a fluctuating
input common to all the oscillators.
The fluctuating input ${\bi I}(t)$ is a piecewise constant random
signal that takes one of $M$ values ${\bi I}_m \in \left\{ {\bi I}_1,
  \cdots, {\bi I}_M \right\}$ with equal probability.
The changes of ${\bi I}(t)$ occur at time $\left\{t_1, t_2, \cdots
\right\}$ following a Poisson process of mean interval $\tau$. We
assume $\tau$ to be sufficiently larger than the period of the
oscillator.
At each $t_{n}$, ${\bi I}(t)$ changes its value in a stepwise manner.
Namely, if ${\bi I}(t_n - 0) = {\bi I}_m$, its new value ${\bi I}(t_n
+ 0)$ after the change is either of ${\bi I}_{m+1}$ or ${\bi I}_{m-1}$
with equal probability.
The probability density function (PDF) of the interval $T_{n} =
t_{n+1} - t_{n}$ between changes obeys an exponential distribution $
P(T) = \exp \left( - T / \tau \right) / \tau$.
For each value of ${\bi I}_m$, Eq.(\ref{Eq:ode}) is assumed to have a
stable limit-cycle solution, whose basin of attraction is the entire
phase space except some unstable fixed points.
\begin{wrapfigure}[16]{r}[0pt]{0.4\linewidth}
  \begin{center}
    \includegraphics[width=0.95\hsize,clip]{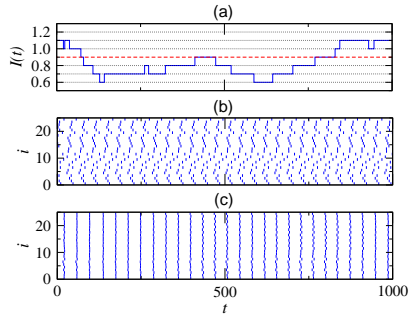}
  \end{center}
  \caption{(a) Typical realization of a piecewise-constant random
    signal. (b) Zero-crossing events under a constant input. (c)
    Zero-crossing events under a fluctuating input.}
  \label{Fig:fig1}
\end{wrapfigure}

Though our theory itself is a general one, we use the FitzHugh-Nagumo
model as an example, where ${\bi X} = \left\{ u, v \right\}$, ${\bi
  G}({\bi X}) = \left\{ \epsilon ( v + a - b u ), v - v^3 / 3 - u
\right\}$, and ${\bi I}(t) = \left\{ 0, I(t) \right\}$.
The parameters are fixed at $a=0.7$, $b=0.8$, and $\epsilon=0.08$.
$I(t)$ takes one of $M=7$ values $I_m \in \{0.6$, $0.7$, $0.8$, $0.9$,
$1.0$, $1.1$, $1.2\}$. We set the mean interval between changes at
$\tau = 40$ and consider $N=25$ oscillators.
In the numerical simulation, small Gaussian-white noise of zero-mean
and intensity $D = 10^{-5}$ is independently applied to each variables
of the oscillators to incorporate the effect of external disturbances.
Figure~\ref{Fig:fig1}(a) displays a typical realization of the
piecewise-constant signal, Fig.~\ref{Fig:fig1}(b) zero-crossing events
of the $v$-component from $v<0$ to $v>0$ under the constant input, and
Fig.~\ref{Fig:fig1}(c) zero-crossing events under the fluctuating
input, sufficiently after initial transients. Due to the independent
Gaussian-white noises, the zero-crossing events occur randomly under
the constant input as shown in Fig.~\ref{Fig:fig1}(b), whereas phase
synchronization induced by fluctuating input can clearly be seen in
Fig.~\ref{Fig:fig1}(c).

\section{Reduction to random phase maps}

The phase synchronization is the result of the stabilization of each
limit-cycle oscillator against phase disturbances due to the
fluctuating input. To analyze its mechanism, we reduce our model to
random phase maps.
We consider the single-oscillator problem, because the stability is a
property of individual oscillators.

\begin{wrapfigure}[15]{r}[0pt]{0.4\linewidth}
  \begin{center}
    \includegraphics[width=0.9\hsize,clip]{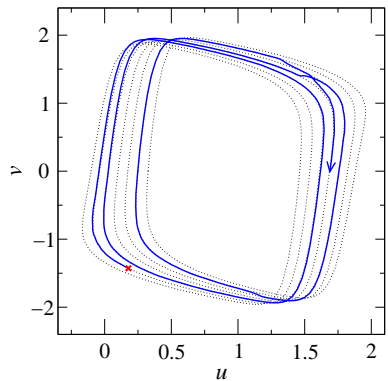}
  \end{center}
  \caption{Typical trajectory of the model (bold arrow). $M$
    limit-cycle orbits corresponding to the $M$ input values are shown
    in the background.}
  \label{Fig:fig2}
\end{wrapfigure}

Corresponding to the $M$ values of ${\bi I}(t)$, the orbit of our
model moves among $M$ limit cycles. Since $\tau$ is assumed to be
large, the orbit is on one of those limit cycles most of the time,
except for short transients between limit cycles after the changes of
the input, as shown in Fig.~\ref{Fig:fig2}.
Following the standard procedure~\cite{Winfree,Kuramoto}, we define a
phase variable $\theta_{m}({\bi X}) \in [0,1]$ using the limit cycle
$m$ corresponding to the input ${\bi I}_{m}$ for each $m = 1, \cdots,
M$, where $0$ and $1$ represent the same phase.
We specify the value of ${\bi I(t)}$ by $m$ hereafter. When the input
is $m$, i.e., ${\bi I}(t) = {\bi I}_m$, the dynamics of the orbit can
simply be described as $ \dot{\theta}_{m}(t) = \omega_m$ by using the
corresponding phase variable $\theta_{m}$, where $\omega_m$ is the
angular velocity of the limit cycle $m$.

When the input changes from $m$ to $m'$, the orbit of our model
originally at phase $\theta_{m}$ on the limit cycle $m$ will be mapped
to new phase $\theta_{m'}$ on the limit cycle $m'$. We describe this
mapping by $ \theta_{m'} = F_{m \to m'}(\theta_m) $, which we call a
``phase map''. It is a periodic function on $[0,1]$ satisfying $F_{m
  \to m'}(\theta_m + 1) = F_{m \to m'}(\theta_m) + 1 = F_{m \to
  m'}(\theta_m)$, where $0$ and $1$ should be interpreted as the same
phase.
Figure ~\ref{Fig:fig3} displays the phase maps of the FitzHugh-Nagumo
model obtained for all contiguous pairs of $(m, m')$. The curves are
appropriately shifted to adjust their origins.

We use the time step $n$ rather than the real time $t$ in the
following discussion, which is the number of changes in ${\bi I}(t)$
from the beginning. Since we consider a Poisson process, the time step
$n$ roughly corresponds to the real time $t$ as $n \simeq t / \tau$,
because the mean inter-impulse interval is $\tau$.
Let us represent the temporal sequence of ${\bi I}(t)$ by $m(n)$, and
consider a situation where the input changes from $m(n)$ to a new
value $m(n+1)$ at $t = t_{n}$ and keeps this value until $t = t_{n+1}$
for an interval of $T_{n} = t_{n+1} - t_{n}$.
The corresponding dynamics of the orbit from phase $\theta_{m(n)}(n)$
on the limit cycle $m(n)$ to the new phase $\theta_{m(n+1)}(n+1)$ on
the limit cycle $m(n+1)$ can be described using the phase map $F$ as
\begin{equation}
  \theta_{m(n+1)}(n+1) = \omega_{m(n+1)} T_{n} + F_{m(n) \to m(n+1)}(\theta_{m(n)}(n)),
  \label{Eq:randommaps}
\end{equation}
where $\omega_{m(n+1)} T_{n}$ represents constant increase of the
phase on the limit cycle $m(n+1)$. Since $T_{n}$ is a random variable,
this equation describes random phase maps.

\section{Stability against phase disturbances}

The stability against phase disturbances can be characterized by the
average Lyapunov exponent of the random phase maps,
Eq.~(\ref{Eq:randommaps}).
Let us consider a small phase deviation $\Delta \theta_{m(n)}(n)$ from
$\theta_{m(n)}(n)$. Its linearized evolution equation is
\begin{equation}
  \Delta \theta_{m(n+1)}(n+1) =
  F'_{m(n) \to m(n+1)}(\theta_{m(n)}(n)) \Delta \theta_{m(n)}(n),
\end{equation}
where $F'_{m \to m'}(\theta_{m}) = d F_{m \to m'}(\theta_{m}) / d
\theta_{m}$. Therefore, the phase deviation grows as
\begin{equation}
  \left| \Delta \theta_{m(n)}(n) / \Delta \theta_{m(0)}(0) \right|
  =
  \prod_{n'=0}^{n-1} \left| F'_{m(n') \to m(n'+1)}(\theta_{m(n')}(n')) \right|
  \simeq
  \exp \left( \lambda n \right),
\end{equation}
where we defined the average Lyapunov exponent as $ \lambda = \langle
\log \left| F'_{m \to m'}(\theta_{m}) \right| \rangle $. The average
should be taken over all possibilities of $(m, m')$ and over the phase
distributions on all limit cycles.

When the mean interval $\tau$ is sufficiently large, the phase
distribution on each limit cycle tends to be uniform, because the
jumps between the limit cycles occur irrespectively of where the orbit
is, leading to complete randomization of the phase.
Under this condition, we can make a general statement on the
sufficient condition for the phase synchronization: {\it when all
  phase maps $F_{m \to m'}(\theta_{m})$ are monotonically increasing
  non-identity functions, the Lyapunov exponent $\lambda$ is negative,
  leading to fluctuation-induced phase synchronization}.

\begin{wrapfigure}[11]{r}[0pt]{0.4\linewidth}
  \begin{center}
    \includegraphics[width=0.95\hsize,clip]{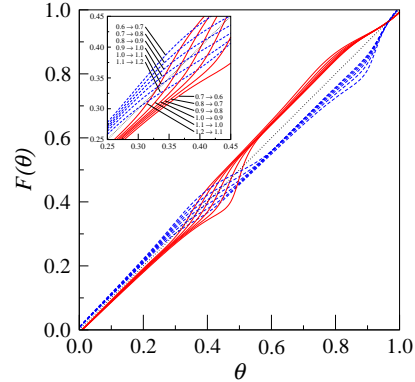}
  \end{center}
  \caption{Phase maps $F_{m \to m'}(\theta)$ between two successive
    values of the input values $m$ and $m'$. The inset is an
    enlargement showing the correspondence between the curve and the
    pair of input values $(m, m')$.}
  \label{Fig:fig3}
\end{wrapfigure}
Actually, when $F'_{m \to m'}(\theta_{m}) > 0$ holds for all $m$, we
can bound the Lyapunov exponent $\lambda$ from above as
\begin{eqnarray}
    \lambda
    &=&
    \frac{1}{\#(m,m')} \sum_{(m,m')}
    \int_{0}^{1} d\theta_m \log F'_{m \to m'}(\theta_{m})
    \cr
    &\leq&
    \frac{1}{\#(m,m')} \sum_{(m,m')}
    \int_{0}^{1} d\theta_m \left\{ F'_{m \to m'}(\theta_{m}) - 1 \right\}
    \cr
    &=&
    \frac{1}{\#(m,m')} \sum_{(m,m')}
    \left\{
      \left[ F_{m \to m'}(\theta_{m}) \right]_{0}^{1} - 1
    \right\}
    =
    0,
    \cr
    &&
  \end{eqnarray}
where the summation is taken over all combinations of $m$ and $m'$,
and $\#(m,m')$ represents the number of them. In the above
inequalities, we utilized the fact that $\log F' \leq F' - 1$, and
that $F_{m \to m'}(1) - F_{m \to m'}(0) = 1$ because $F_{m \to
  m'}(\theta_{m})$ is a phase map. The equality holds only when $F'_{m
  \to m'}(\theta_{m}) \equiv 1$ for all $m$, namely, when the phase
maps are trivial identity maps.
For the FitzHugh-Nagumo model with the parameter values assumed here,
all the phase maps $F_{m \to m'}(\theta_{m})$ are monotonically
increasing as can immediately be seen from Fig.~\ref{Fig:fig3}.
Therefore, by applying a piecewise-constant random signal with large
mean interval $\tau$, fluctuation-induced synchronization occurs as
demonstrated in Fig.~\ref{Fig:fig2}(c). In general, as long as the
separation between neighboring values of ${\bi I}(t)$ are small, the
phase maps should be monotonic, and fluctuation-induced
synchronization should occur.

\section{Summary}

We analyzed fluctuation-induced phase synchronization among uncoupled
noisy oscillators for the case of a slowly varying, piecewise-constant
random input.  By reducing the model to random phase maps, we gave a
general sufficient condition for the phase synchronization. Extension
of our current analysis to a realistic continuous random signal will
be tackled in the future.

\section*{Acknowledgments}

H. N. is deeply indebted to Professor Yoshiki Kuramoto for his
continuous support. We also thank Y. Tsubo. D. Tanaka, and J.  Teramae
for useful comments.

\end{document}